# Generic Cryogenic CMOS Device Modeling and EDA-Compatible Platform for Reliable Cryogenic IC Design

Zhidong Tang[1,5#], Zewei Wang[1,2#], Yumeng Yuan[1,2#], Chang He[1], Xin Luo[3], Ao Guo[3], Renhe Chen[1], Yongqi Hu[1,2], Longfei Yang[3], Chengwei Cao[3], Linlin Liu[3], Liujiang Yu[4], Ganbing Shang[4], Yongfeng Cao[4], Shoumian Chen[3], Yuhang Zhao[3], Shaojian Hu[3], and Xufeng Kou[1*](Senior Member, IEEE)

[1]ShanghaiTech University, Shanghai, China, [2]University of Chinese Academy of Sciences, Beijing, China, [3]Shanghai IC Research and Development Center, Shanghai, China, [4]Huali Microelectronics Corporation (HLMC), Shanghai, China, [5]School of Integrated Circuits, Tsinghua University, Beijing, China.
*Email: kouxf@shanghaitech.edu.cn; [#]Authors contributed equally to this work.

*Abstract*—This paper outlines the establishment of a generic cryogenic CMOS database in which key electrical parameters and transfer characteristics of the MOSFETs are quantified as functions of device size, temperature/frequency responses. Meanwhile, comprehensive device statistical study is conducted to evaluate the influence of variation and mismatch effects at low temperatures. Furthermore, by incorporating the Cryo-CMOS compact model into the process design kit (PDK), the cryogenic 4 Kb SRAM, 5-bit flash ADC and 8-bit current steering DAC are designed, and their performance is readily investigated and optimized on the EDA-compatible platform, hence laying a solid foundation for large-scale cryogenic IC design.

## I. INTRODUCTION

Cryogenic CMOS (Cryo-CMOS) has shown great potential for enabling versatile energy-efficient computing paradigms in the upcoming post-Moore era. Due to the inherent improvements of MOSFET device performance at low temperatures in terms of enhanced driving strength, smaller subthreshold swing, diminished leakage current, and lower interconnect resistance, a fully Cryo-CMOS-enabled digital computer is projected to achieve a 3.4× higher processing speed and or 37% reduction in power consumption when operated at 77 K compared to the room-temperature performance [1]. In the meantime, a 6T-SRAM bit-cell, which is re-designed through the design-technology co-optimization (DTCO) for 77 K operation, can boost the speed by 1.7× at a fixed power budget (or save 80% power consumption at the same frequency) [2]. Moreover, recent advancements in quantum computing have further broadened the research scope of Cryo-CMOS. For instance, the scalable quantum processor architecture requires the placement of the control, amplification, and readout modules adjacent to the qubit gate arrays so that the thermal noise and communication delay are minimized [3][4]. Accordingly, various Cryo-CMOS-based digital/analog integrated circuits (*e.g.*, LNA, PLL, DAC, and ADC) have been developed to bridge the qubits and the user interface [5]-[7]. In addition, cryogenic electronics also play an important role in aerospace exploration, medical and scientific applications, in which they can facilitate the design of low-noise sensors and high-precision controllers [8]-[10].

In order to meet the growing demand of these emerging applications, it is of great importance to establish a generic platform that guides reliable and efficient cryogenic integrated circuit designs [11]. In particular, such a Cryo-CMOS platform should include sufficient characterizations of both transistors and back-end-of-line (BEOL) components under low temperatures. Concurrently, the relevant device compact models and process design kits (PDK) should maintain a high accuracy in a wide temperature range. In this regard, while low-temperature device physics has been investigated and relevant Cryo-CMOS device models have been developed for various CMOS technology processes [12], yet the application of these cryogenic databases at the circuit/system level is still in its early development stage (*e.g.*, most EDA tools still do not support circuit simulations below 230 K) [1].

In this work, we present the establishment of a generic design platform for Cryo-CMOS applications. The cryogenic modeling flow includes the golden die selection, temperature-dependent device DC/RF characterizations, key electrical parameter extraction, and statistical device variation analysis, hence capturing the device electrical characteristics with varied

Manuscript received February 8, 2024. This work was supported by National Key R&D Program of China (2021YFA0715503, 2023YFB4404000), National Natural Science Foundation of China (92164104), the Strategic Priority Research Program of CAS (XDA18010000), Shanghai Rising-Star Program (21QA1406000) and the Open Fund of State Key Laboratory of Infrared Physics. Z. Tang, Z.W Wang and Y. Yuan contribute equally to this work.

Z. Tang, Z. Wang, Y. Yuan, C. He, R Chen, Y. Hu, and X. Kou are with School of Information Science and Technology, ShanghaiTech University, Shanghai 201210, China (e-mail: kouxf@shanghaitech.edu.cn).

Z. Wang, Y. Yuan, Y. Hu are also with Shanghai Institute of Microsystem and Information Technology, Chinese Academy of Sciences, Shanghai 200050, China.

X. Luo, A. Guo, L. Yang, C. Cao, L. Liu, S. Chen, Y. Zhao, and S. Hu are with the Shanghai IC Research and Development Center (ICRD), Shanghai 201210, China.

L. Yu, G. Shang, and Y. Cao are with Huali Microelectronics Corporation (HLMC), Shanghai 201314, China.

Z. Tang is also with the School of Integrated Circuits, Beijing Advanced Innovation Center for Integrated Circuits, Tsinghua University, Beijing 100084, China.

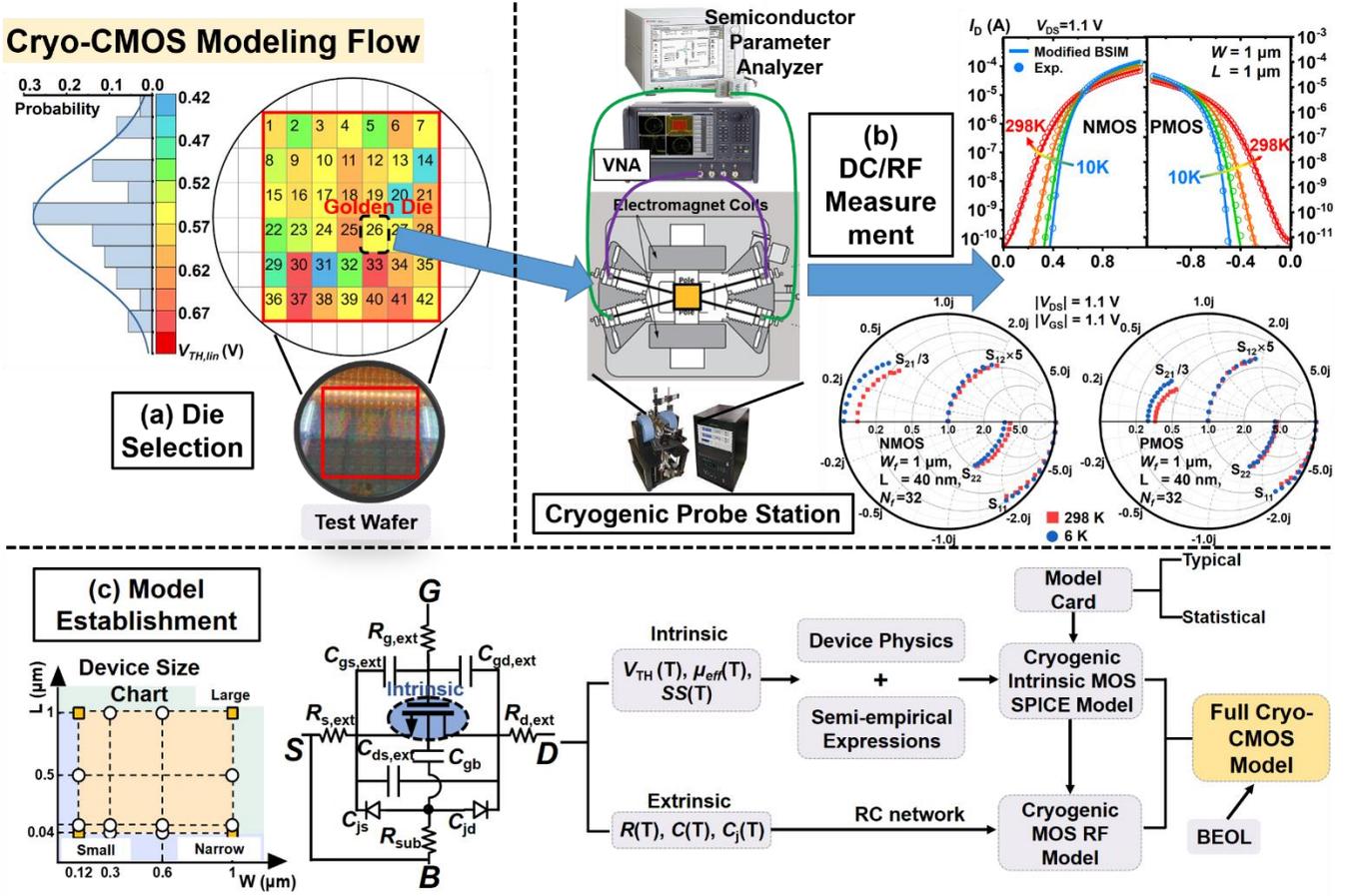

**Fig. 1.** Illustration of cryogenic CMOS modeling flow. (a) Wafer mapping for golden die selection. (b) Temperature-dependent DC/RF device characterizations. (c) Generic Cryo-CMOS device compact model establishment.

gate geometries, temperatures, and frequencies. Moreover, by integration the modified Cryo-BSIM compact model into the PDK library, we demonstrate the design of three cryogenic digital (SRAM) and analog (ADC/DAC) circuit modules which leverage the device-level operating mechanisms to achieve an optimized circuit performance at low temperatures.

## II. CRYOGENIC CMOS DEVICE MODELING

### A. Cryogenic Device Modeling Flow

Fig. 1 outlines the general Cryo-CMOS device characterization and modeling flow adopted in this study. With the help of the Automatic Test Equipment (ATE), the golden test die (*i.e.*, whose electrical benchmarks are close to the process statistical average specifications at room temperature) was firstly selected from 42 Dies Under Test (DUTs) on the wafer of the HLMC 40-nm technology node (Fig. 1a). Equipped with Lakeshore cryogenic probe station, the high-precision Keysight B1500A semiconductor device analyzer and the Keysight 5227A Vector Network Analyzer (VNA), we have developed a suitable cryogenic test platform which can cover wide temperature (10 K – 300 K) and frequency (250 MHz – 40 GHz) ranges. Accordingly, systematic device DC/RF measurements with respect to different temperatures and device sizes were carried out to ensure the full-scale Cryo-CMOS device modeling (Fig. 1b).

Based on the experimental dataset and low-temperature device physics, relevant cryogenic device SPICE models with semi-empirical formulas were subsequently generalized so that key electrical parameters (*e.g.*, threshold voltage, channel mobility, and sub-threshold swing) in the entire examined temperature/size region can be accurately described by one set of fitting parameters (*i.e.*, model card). Concurrently, considering that the layout-dependent high-frequency RC effects become indispensable for analog and RF IC design, we also calibrated the temperature dependence of extrinsic resistive/capacitive/inductive components in reference to the equivalent small-signal MOSFET circuit model. Besides, in order to evaluate the impact of manufacturing process fluctuations on device performance at cryogenic temperatures, a statistical model card, which consists of a set of random variables used for Monte-Carlo simulations, was developed based on the 42 DUTs data.

In addition to the front-end-of-line (FEOL) device study, the back-end-of-line (BEOL) components (*e.g.*, resistors, capacitors, and interconnect) were characterized in the cryogenic temperature region to account for the parasitic effects on cryogenic integrated circuit simulations. Afterwards, the SPICE models of transistors and relevant BEOL components were incorporated into the generic cryogenic CMOS PDK library. As a result, our established Cryo-CMOS platform could capture the temperature/frequency/geometric response of MOSFETs and extend the EDA simulation tool down to 10 K, hence providing a reliable guidance for the large-scale IC design. In the following sub-sections, cryogenic MOSEFTs device modeling, statistical variation/mismatch analysis, as well as model validation will be discussed in detail.

*B. Cryogenic Intrinsic MOSFET Device Modeling*

In the standard EDA simulation flow, model cards (*i.e.*, which are provided by foundries and include the technology process information such as oxide thickness, doping profile, threshold voltage, and channel mobility) and netlists (*i.e.*, which list the components in the circuit and the node connection information) are functioned as input parameters in the model simulator framework, and the output node information includes voltage, current, and capacitance. In this context, the accuracy of the EDA simulator not only depends on the model cards, but also relies on the device compact model framework that can describe the electrical characteristics of MOSFETs. However, most commercial models only support the temperature range of [233 K, 393 K], and they do not consider cryogenic device physics such as the Fermi level shift, phonon/ionized impurity scatterings, carrier freeze-out, and band tail effect [12]-[15].

To address such challenge, we have performed temperature-dependent *I-V* characterizations of MOSFETs and confirmed that all nano-scale transistors would experience an increased threshold voltage ($V_{TH}$), enhanced channel mobility ($\mu_{eff}$), and steeper sub-threshold swing (SS) when the base temperature gradually drops to cryogenic temperatures. Based on the low-temperature device physics, we have quantified the correlations between these electrical parameters and temperature. Subsequently, we have proposed a modified Cryo-BSIM model with three major changes including $V_{TH}(T)$, $\mu_{eff}(T)$, and $T_{eff}$ in the core BSIM-4 model, as re-captured as follows [17]:

1) In general, the change of bulk carrier density (*i.e.*, which causes the Fermi potential shift) and the broadened depletion width (*i.e.*, which tailors the effective width and length of the inversion channel) lead to an enlarged $V_{TH}$ at low cryogenic temperatures [18]. Accordingly, we have introduced the temperature-driven gate geometry effects in the threshold voltage equation [16]. Besides, the carrier freeze-out effect also changes the substrate resistance and leads to the re-distribution of the electric field in the bulk region, therefore bringing about a modified body effect-related term $\Delta V_{BS}(T)$. Therefore, the overall threshold voltage correction term $\Delta V_{TH}$ is given by

$$\Delta V_{TH}(T) = Z \cdot \Delta V_{TH,W}(T) + K \cdot \Delta V_{TH,L}(T) + \Delta V_{BS}(T) \quad (1)$$

where $\Delta V_{TH,W}(T)$ and $\Delta V_{TH,L}(T)$ correspond to the narrow-width and short-channel effects, and the corresponding fitting parameters $Z$ and $K$ denote their contributions, respectively.

2) The drain current in the MOSFET device is closely associated with the effective channel mobility $\mu_{eff}$ [19]. Given that both the ionization scattering and size limiting factor (*i.e.*, $\mu_{eff} \propto 1/(1+2\lambda_0/L)$, where $\lambda_0$ is the low-field mean-free path) would become the dominant factors at low temperatures, we have proposed a semi-empirical effective mobility model [17]

$$\mu_{eff}(T) = \frac{\mu_0(T/298)^{-1.2}}{f_1(\lambda_0,L) \cdot f_2(T,E_{eff},\theta_1,\theta_2)} \quad (2)$$

where $\mu_0$ is the long-channel low-field mobility at room temperature, functions $f_1$ and $f_2$ represent the temperature-dependent scatterings and dimension-induced ballistic transport restriction effect, $\lambda_0$ is the low-field mean free path, $E_{eff}$ is the effective field associated with the surface roughness scattering, and $\theta_1$ and $\theta_1$ are size-dependent fitting parameters.

3) Instead of a simple linear correlation with the base temperature, the subthreshold swing (SS) is found to exhibit a saturation behavior at deep cryogenic temperatures (*e.g.*, $T < 40$ K in our 40 nm-node devices), possibly due to the disorder-induced band tail broadening and the presence of localized interface states near the band edge [20]. In this regard, we have introduced an effective temperature parameter $T_{eff}$ in the SS model as:

$$T_{eff} = \frac{T+T_0+\sqrt{(T-T_0)^2+D}}{2} \quad (3)$$

where $T_0$ is the critical temperature at which the impact of band-tail occurs, and $D$ is the smoothing parameter.

In conclusion, with the aforementioned modifications based on the original BSIM-4 framework, the updated key parameters in the model card were subsequently fitted in reference to the experimental data. As a result, the modified Cryo-BSIM model can accurately depict the DC transfer characteristics of all transistors across the device size chart on the golden die wafer from room temperature down to cryogenic temperatures, with the average room-mean-square (RMS) values of the fitting error all below 5%, as highlighted in Fig. 3.

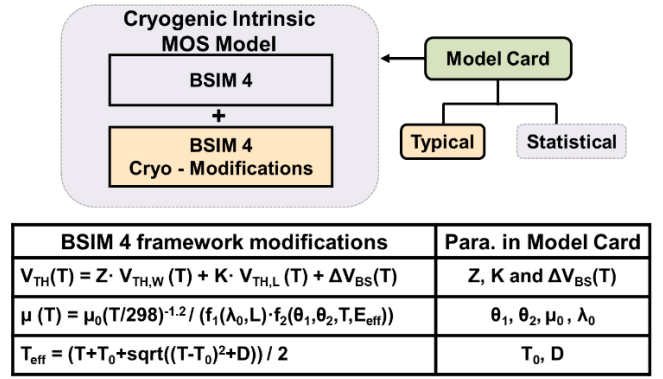

| BSIM 4 framework modifications | Para. in Model Card |
|---|---|
| $V_{TH}(T) = Z \cdot V_{TH,W}(T) + K \cdot V_{TH,L}(T) + \Delta V_{BS}(T)$ | Z, K and $\Delta V_{BS}(T)$ |
| $\mu(T) = \mu_0(T/298)^{-1.2} / (f_1(\lambda_0,L) \cdot f_2(\theta_1,\theta_2,T,E_{eff}))$ | $\theta_1, \theta_2, \mu_0, \lambda_0$ |
| $T_{eff} = (T+T_0+\text{sqrt}((T-T_0)^2+D)) / 2$ | $T_0$, D |

**Fig. 2.** Cryogenic extensions to baseline BSIM-4 model.

*C. Cryogenic MOSFETs Variation Analysis*

In addition to the individual device modeling, it is also critical to include a comprehensive statistical dataset, which accounts for the device performance variations caused by random process fluctuations, in the model card so that the high-precision EDA simulations of the designed integrated circuits are consistent with the tape-out test results [21]. To meet the requirement of a solid statistical analysis, we have further collected the temperature-dependent *I-V* data of MOSFETs from all DUTs with our cryogenic test platform, and generated the statistical model that incorporates both the global variation (*i.e.*, die-to-die and wafer-to-wafer variations) and local mismatch variation (*i.e.*, device-to-device variations on the same die) at cryogenic temperatures, as shown in Fig. 4.

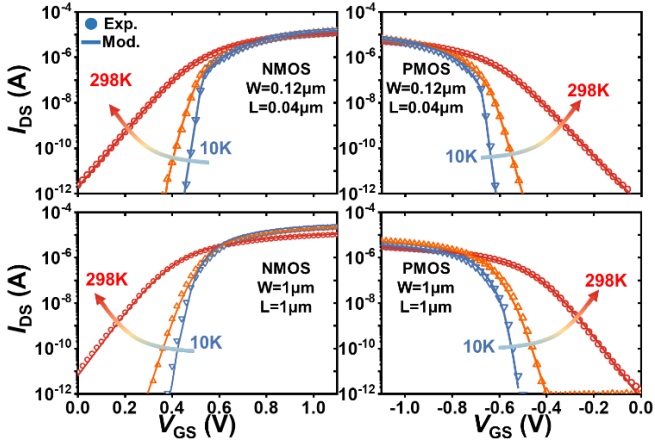

**Fig. 3.** Cryogenic transfer characteristics of NMOS and PMOS devices with validated model fitting.

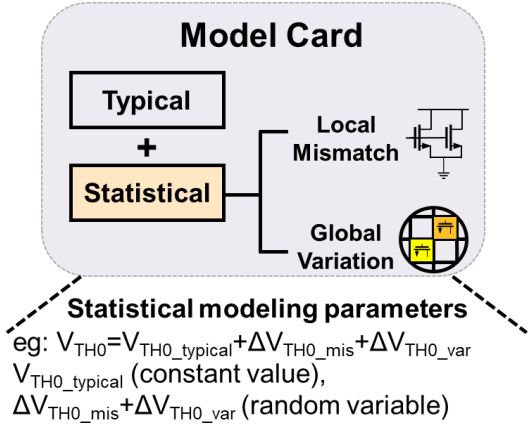

**Fig. 4.** MOSFETs statistical model card.

To exemplify the evolution trend of the global variation versus temperature, we have recorded the $I_{DS}$-$V_{GS}$ results of the MOSFETs from 42 DUTs across the whole wafer. As depicted in Fig. 5, the transfer characteristics curves do display disparities among devices, and such variations become more pronounced in small-size transistors at low temperatures.

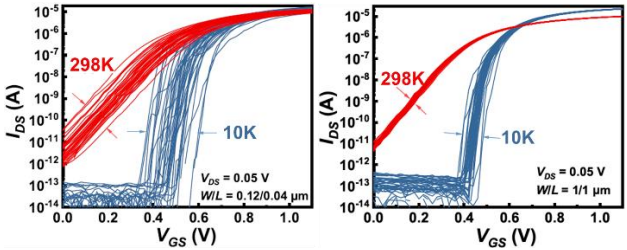

**Fig. 5.** Transfer characteristics of the NMOS devices with (a) $(W, L) = (0.12$ μm, $0.04$ μm$)$ and (b) $(W, L) = (1$ μm, $1$ μm$)$ at $T = 10$ K and 298 K from 42 test dies.

Quantitatively, Fig. 6 summarizes the probability density function (PDF) of the threshold voltage of NMOS devices at 4 size corners (*i.e.*, $W/L = 0.12$ μm/$0.04$ μm (small device), 1 μm/$0.04$ μm (short device), $0.12$ μm/1 μm (narrow device), and 1 μm/1 μm (big device)). It is clear that the global variation $\sigma(V_{TH})$ (*i.e.*, which is defined as the standard deviation of the PDF curve) progressively enlarges as the temperature decreases, and the PDF statistical distribution curves become more broadened for the short-channel devices. By further counting 242 devices in total, the overall global threshold variation of $\Delta V_{TH} = V_{TH} - \overline{V_{TH}}$ (*i.e.*, where $\overline{V_{TH}}$ is the average value of $V_{TH}$) increases by 10% when the base temperature drops from 298 K to 10 K, as visualized in Fig. 7.

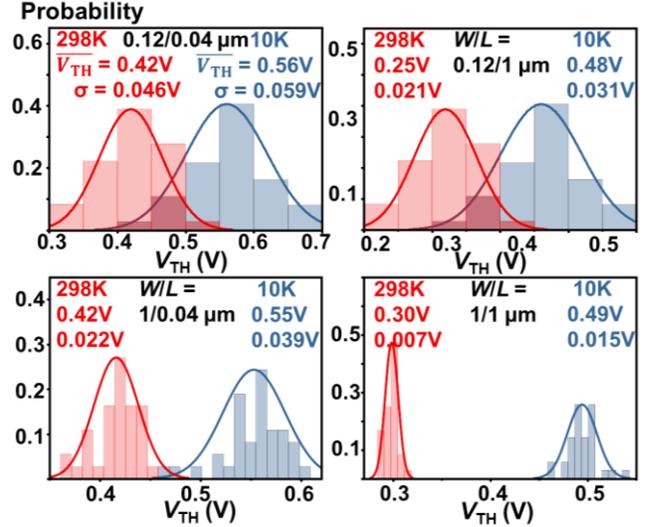

**Fig. 6.** Probability density function of the threshold voltage of NMOS devices with four different gate geometries at $T = 298$ K (red) and 10 K (blue), respectively.

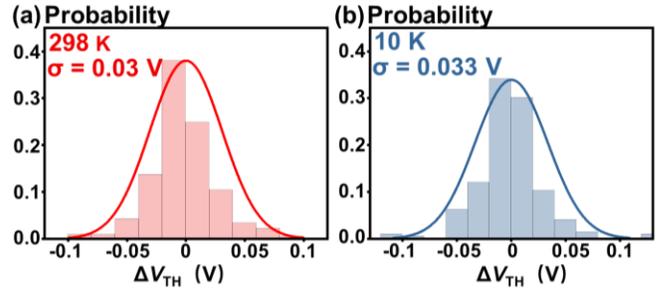

**Fig. 7.** Global variation statistics of $V_{TH}$ counted from a total of 242 NMOS devices at (a) $T = 298$ K and (b) $T = 10$ K.

Moreover, since the differential pair structure has been widely used in analog and RF circuits, any mismatch of the paired transistors caused by local stochastic fluctuations would degrade the circuit performance [22]. Therefore, to examine the local MOSFETs mismatch scenario, the $V_{TH}$ values were extracted from 50 transistors pairs on the same batch of the test-key, with varied gate aspect ratios ranging from $W/L = 0.12$ μm/$0.04$ μm to 1 μm/1 μm. As displayed in Fig. 8, both the $T = 10$ K and 298 K data follow the Pelgrom's law of $\sigma(\Delta V_{TH}) = $ $A_{VTH}/\sqrt{W \cdot L}$ (*i.e.*, where the slop $A_{VTH}$ represents the threshold voltage factor depending on the fabrication process) [23], yet the characteristic value of $A_{VTH}$ increases by 78% when the transistors are cooled down from 298 K to 10 K, which indicates the worsened fluctuations of the doping profile at lower temperatures.

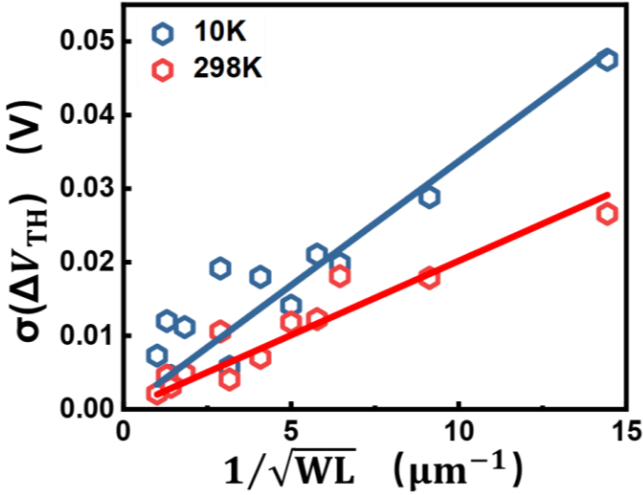

**Fig. 8.** Pelgrom's plots of the fitted $\sigma(\Delta V_{TH})$ lines in reference to the experiment data at $T$ = 298 K and 10 K.

Based on these quantitative variation and mismatch results, we have completed the statistical model card files with additional temperature-dependent random variables in the Cryo-BSIM framework. For example, we have introduced both the calibrated $\sigma(T)$ and $A_{VTH}(T)$ coefficients into the modified low-temperature threshold voltage variation model as

$$V_{TH0}= V_{TH0\_typical} + (1+\sigma(T)) \cdot dV_{TH0\_mis} + (1+ A_{VTH}(T)) \cdot dV_{TH0\_var} \quad (4)$$

where $V_{TH0\_typical}$ is the reference mean value of the golden-die device, while $dV_{TH0\_mis}$ and $dV_{TH0\_var}$ are the local mismatch and global variation-associated random variables which both follow the normal distribution relations. Consequently, the updated cryogenic PDK library is able to effectively cover the 10 K ≤ $T$ ≤ 298 K range. Fig. 9 illustrates the tracking plots from 1000 times overall Monte-Carlo simulations (*i.e.*, which include both the local mismatch and global variation). As a summary, the statistical study in this sub-section provides a trustworthy direction to mitigate the influence of the manufacturing process-induced variation and mismatch effects by adopting appropriate device size engineering during the cryogenic circuit and system design.

*D. Cryogenic MOSFETs High-Frequency Device Modeling*

In order to achieve a reliable RF modeling of the Cryo-CMOS devices, we have carried out temperature-dependent S-parameter characterizations (from 0.25 to 40 GHz) on 36 multi-finger gate NMOS and PMOS devices across the device size chart of the HLMC 40-nm RF test wafer [24]. As shown in Fig. 10, the equivalent small-signal equivalent circuit of the RF Cryo-CMOS device consists of the intrinsic MOSFET model elaborated in Section II.B (*i.e.*, which governs the DC electrical properties), and the extrinsic *RC*-network associated with the ground-signal-ground (GSG) device layout structure (*i.e.*, which determines the input/output impedance, pole/zero positions, as well as affects the high-frequency figure-of-merit of the designed circuit). Afterwards, the measured S-parameter data were transformed into the Y- and Z-parameter matrices, and the conventional Cold-FET method was adopted to extract the corresponding layout-dependent resistance and capacitance values. Meanwhile, temperature-dependent Short-Open-Load-Through (SOLT) method (*i.e.*, by using the standard GGBCS-5 calibration substrate) and Open-Short de-embedding technique were also conducted to exclude parasitic components from the instrument, cables, and on-chip interconnects/GSG pads [25]. Besides, we need to point out that the passive inductive components do not show any temperature dependence, and we have hence treated them as a set of constants in the cryogenic RF device model.

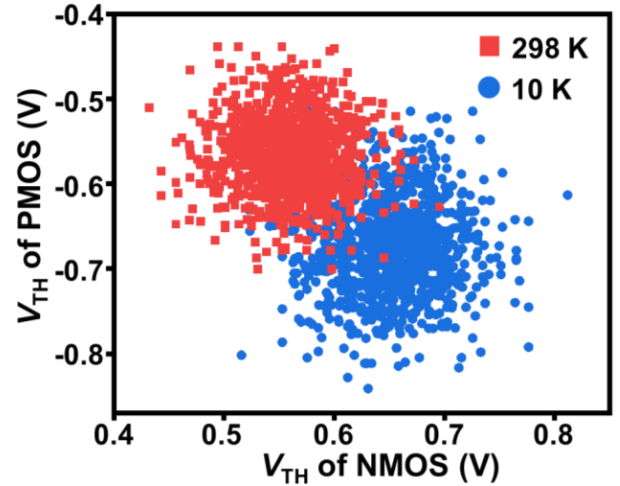

**Fig. 9.** Tracking plot of the $V_{TH}$ from 1000 times Monte-Carlo simulations at $T$ = 298 K (red) and 10 K (blue). The size of the transistor used in this figure is chosen as $W/L$ = 0.12 μm/0.04 μm.

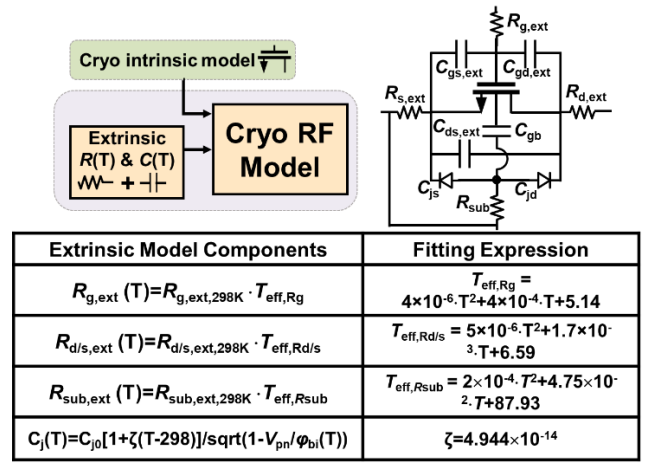

| Extrinsic Model Components | Fitting Expression |
|---|---|
| $R_{g,ext}(T)=R_{g,ext,298K} \cdot T_{eff,Rg}$ | $T_{eff,Rg}$ = $4 \times 10^{-6} \cdot T^2 + 4 \times 10^{-4} \cdot T + 5.14$ |
| $R_{d/s,ext}(T)=R_{d/s,ext,298K} \cdot T_{eff,Rd/s}$ | $T_{eff,Rd/s}$ = $5 \times 10^{-6} \cdot T^2 + 1.7 \times 10^{-3} \cdot T + 6.59$ |
| $R_{sub,ext}(T)=R_{sub,ext,298K} \cdot T_{eff,Rsub}$ | $T_{eff,Rsub}$ = $2 \times 10^{-4} \cdot T^2 + 4.75 \times 10^{-2} \cdot T + 87.93$ |
| $C_j(T)=C_{j0}[1+\zeta(T-298)]/\sqrt{1-V_{pn}/\varphi_{bi}(T)}$ | $\zeta = 4.944 \times 10^{-14}$ |

**Fig. 10.** Equivalent RF circuit of the Cryo-CMOS device which includes the intrinsic MOSFET DC model and the layout-dependent resistive and capacitive components.

Fig. 11 summarizes the changes of extrinsic resistances and capacitances as functions of temperature and frequency. Owning to the degenerately-doped poly-silicon gate and silicide Source/Drain area, both the gate electrode resistance $R_{g,ext}$ and Source/Drain terminal resistance $R_{s/d,ext}$ become smaller with the decrease of temperature, and such metallic-like $R$-$T$ behaviors can be described by the 2nd-order polynomial equation $R(T) = R(298\ \text{K}) \cdot T_{eff,R}$, where $T_{eff,R} = A \times (T - 298)^2 + B \times (T - 298) + 1$, and the fitting parameters $A$ and $B$ are only dependent on materials [26]. In view of the RF MOSFET-related capacitive components, their temperature-dependent slopes can be clearly divided into two categories. On the one hand, the metal routing capacitances ($C_{gs/gd,ext}$ and $C_{ds,ext}$) stay almost unchanged with temperature, owning to their parallel plate capacitor nature. On the other hand, the diffusion capacitances ($C_{js}$, $C_{jd}$) in the Source/Drain-to-bulk junctions can be modeled as $C_j(T) = C_{j0}[1 + \zeta(T - 298)]/\sqrt{1 - V_{pn}/\varphi_{bi}(T)}$, where $\zeta$ is the fitting parameter affiliated with the temperature correction term, $V_{pn}$ is the applied voltage, and $\varphi_{bi}(T)$ is the built-in potential across the junction region. Therefore, after accomplishing the data fitting procedure (*i.e.*, using the ICCAP software) for the cryogenic model card, the updated Cryo-BSIM library manages to cover the NMOS and PMOS devices in the 0.04 μm ≤ $L$ ≤ 1 μm, 10 K ≤ $T$ ≤ 298 K and 0.25 GHz ≤ $f$ ≤ 40 GHz region.

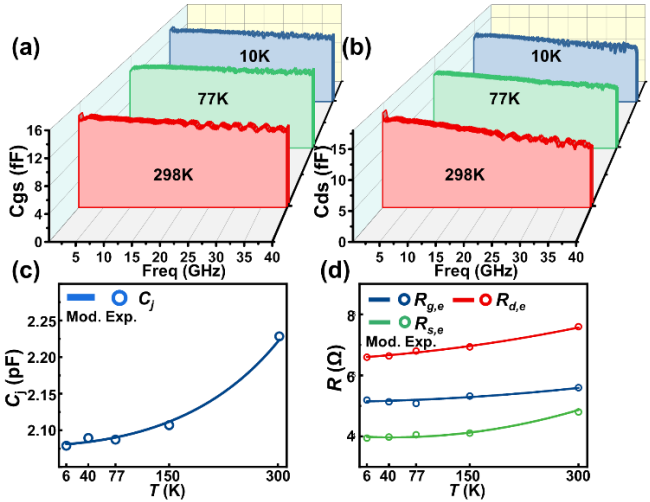

**Fig. 11.** Frequency-dependent (a) $C_{gs}$ and (b) $C_{ds}$ extracted from the S-parameter data with varied temperatures. Temperature-dependent (c) $C_j$ and (d) $R_{g,ext}/R_{s/d,ext}$ of the RF MOSFETs with $W = 1$ μm, $L = 0.04$ μm, and $N_f = 32$.

*E. Cryogenic Back-End-of-Line Device Characterization*

In addition to MOSFETs, the delineations of the BEOL components (*e.g.*, resistors, capacitors, and interconnecting metal layers) are also of great importance for the cryogenic IC design as they determine the interconnect delay, power consumption, channel capacity, and the displacement/routing strategy. Accordingly, Fig. 12 presents the temperature-dependent electrical behaviors of such passive devices. As the base temperature gradually decreases, the resistances of all metallized diffusion and poly-Si types monotonically reduce (Fig. 12a). In stark contrast, their non-metallized counterparts (SAB-type) is less susceptible to the temperature variation, which could be invaluable for cryogenic bias circuit design (Fig. 12b). Meanwhile, all interconnecting metal layers follow a universal metallic $R$-$T$ behavior with a distinct resistance reduction by more than 50% at $T = 10$ K (Fig. 12c). Such a salient feature could considerably enhance the energy efficiency of the system at cryogenic temperatures. Additionally, Fig. 12d reveals that the MOM capacitance of the BEOL technology remains almost constant regardless of the temperature variation, thus providing a quite stable reference for Cryo-CMOS analog and RF circuit design.

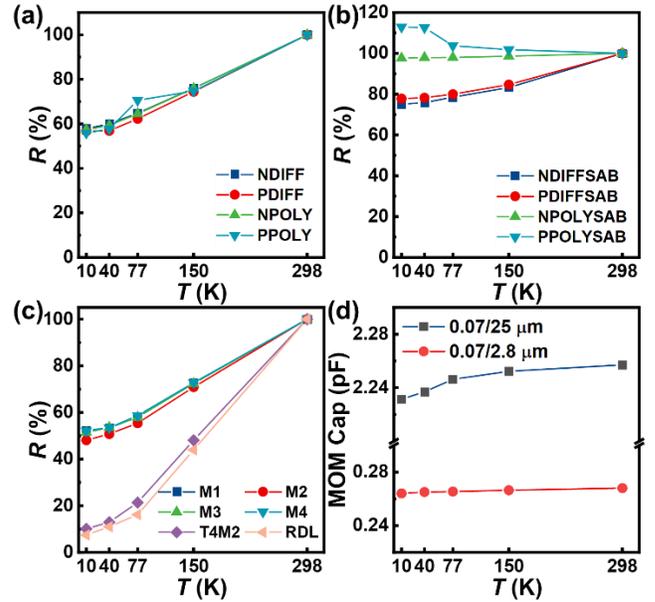

**Fig. 12.** Temperature dependence of the (a) metallized resistors, (b) silicide block type resistors, (c) interconnecting and metal layers, and (d) MOM capacitors.

*F. Validation of the Cryogenic Device Model*

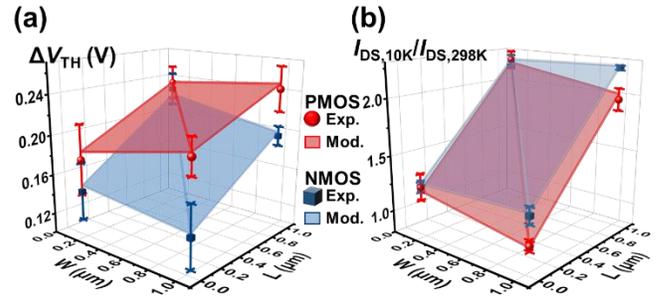

**Fig. 13.** Statistics of (a) $\Delta V_{TH} = V_{TH,10K} - V_{TH,298K}$ and (b) relative $I_{DS,10K}/I_{DS,298K}$ with NMOS and PMOS transistors with four corner sizes. The error bars indicate the variation range of the experiment data collected from 42 test-dies, while the solid dots represent the mean values used in the Cryo-CMOS model.

The validation of the proposed cryogenic device model is presented in this sub-section. First of all, to minimize the device variation/mismatch-induced error, we have assigned the mean values of the threshold voltage ($V_{TH}$) and on-state current ($I_D$) as the reference points during the device modeling process, as highlighted in Fig. 13. Under such circumstances, although the device-to-device variation increases in the low-temperature region (*i.e.*, discussed in Section II.C), the relative root-mean-square deviation of the $I_{DS}$-$V_{GS}$ curves between the Cryo-BSIM compact model and the experimental data only slightly enlarges from $RMS_{298K} < 3\%$ to $RMS_{10K} < 10\%$ at $T = 10$ K, as visualized in Fig. 14a.

Next, given that the intrinsic trans-conductance ($g_m$) and cut-off frequency ($f_T$) can be explicitly expressed as [27]

$$g_m(T) \approx \mu_{\text{eff}}(T) C_{ox} \frac{W}{L} (V_{GS} - V_{TH}(T)) \quad (5)$$

$$f_t(T) = \frac{g_m(T)}{2\pi \left[ C_{gs}\left(1+\frac{R_d(T)+R_S(T)}{r_o}\right) + C_{gd}(1+(R_d(T)+R_S(T))(g_m(T)+1/r_o)) \right]}$$

where $r_0$ is the small-signal output impedance of the device. Accordingly, by successively substituting the temperature-dependent $\mu_{\text{eff}}(T)$, $V_{TH}(T)$, $R(T)$, and $C(T)$ models into the above equations, the resulting simulated $f_T$-$T$ results in Fig. 14b are highly consistent with the experimental values which are directly extracted from the $|H_{21}|$ plots.

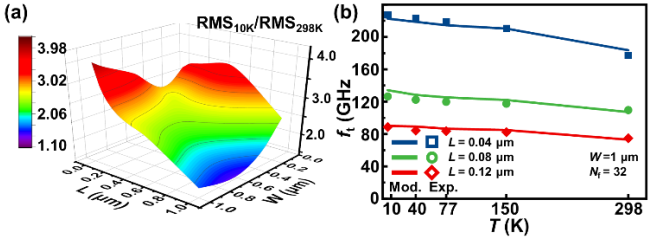

**Fig. 14.** (a) Relative RMS deviation of the $I_{DS}$-$V_{GS}$ transfer characteristics between the Cryo-CMOS compact model and the experimental data. (b) Comparisons of the temperature-dependent cut-off freqency data from simulations and measurements.

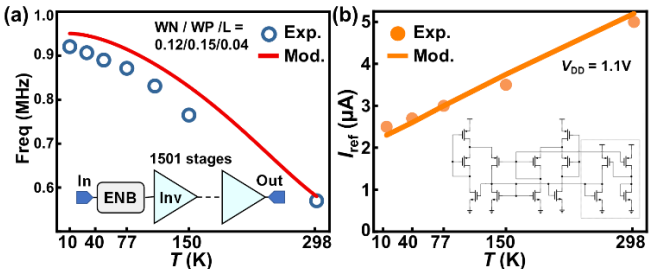

**Fig. 15.** Verification of the taped-out experiment results and the post-layout simulation results of (a) 1501-stage RO and (b) bandgap reference at different temperatures.

Furthermore, the accuracy of our Cryo-CMOS model was also verified at the circuit level. In particular, according to the cryogenic PDK library, we have designed a 1501-stage ring oscillator (RO) and a bandgap reference (BG) which aim to operate in the entire temperature region. Fig. 15 justifies that the post-layout simulation results of these two circuits are in good agreement with the experimental taped-out data from room temperature down to $T = 10$ K, again manifesting the reliability of our cryogenic modeling strategy.

## III. CRYOGENIC CIRCUIT DESIGN BASED ON GENERIC CRYO-CMOS PLATFORM

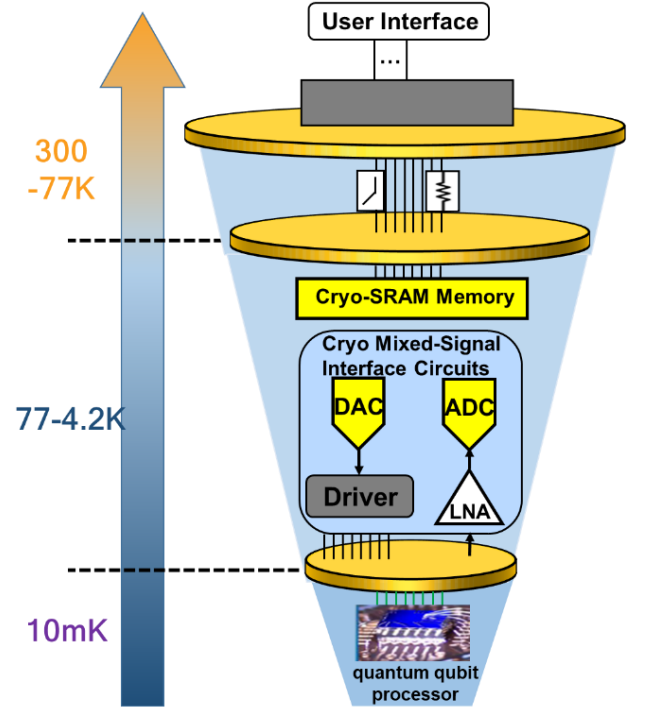

**Fig. 16.** Schematic of the solid state-based quantum computer architecture with integrated quantum control system in a cryogenic environment.

After quantifying the low-temperature electrical properties of the CMOS transistors and BEOL components with the Cryo-CMOS model, we have further applied the modified cryogenic PDK library to guide the design and optimization of various digital and analog integrated circuits for the emergent cryogenic electronic applications. For instance, in a scalable solid state-based quantum computer system illustrated in Fig. 16, the generation of entangled qubits is realized at a deep cryogenic temperature (*e.g.*, $T \approx 10$ mK), and the output signals of the quantum processor are typically in the form of mV-level pulses with the operating frequency in the [1 GHz, 20 GHz] range [28]-[30]. In this context, in order to maintain a high fidelity of the output states, the complementary quantum control system is preferred to be placed at 4.2 K (liquid helium) $\leq T \leq$ 77 K (liquid nitrogen) so that not only the thermal noise is reduced, but also the manipulation, conversion, and error correction of qubits can be finished within the coherent time [31]. Besides, it is also recommended to allocate a Cryo-SRAM memory in the same cryogenic temperature zone so as to minimize the propagation delay between the computing and storage modules.

Following this general architecture, the cryogenic memory module of the quantum computer was firstly designed. Considering the driving strengths of the NMOS and PMOS transistors change with temperature because of the mobility mismatch, we have re-adjusted the PMOS-to-NMOS gate-size ratio of the 6T SRAM bit-cell to optimize its drive capacity and static noise tolerance. Meanwhile, the peripheral circuitries (*e.g.*, pre-charge stage, sense amplifier, and address decoder) have also been modified according to the cryogenic PDK library to ensure their correct functions down to 10 K. Fig. 17a shows the post-layout simulation results regarding the write/read dynamics and power dissipation of the cryogenic 4 Kb SRAM memory. From these data, we can subsequently take full advantage of the EDA tool to facilitate the optimization process. As a result, the final taped-out 4 Kb Cryo-SRAM circuit indeed exhibits better performance in terms of speed and power consumption at low temperatures (Fig. 17b), and its quiescent supply current (IDDQ) benchmark is in good agreement with our model estimation (Fig. 17c).

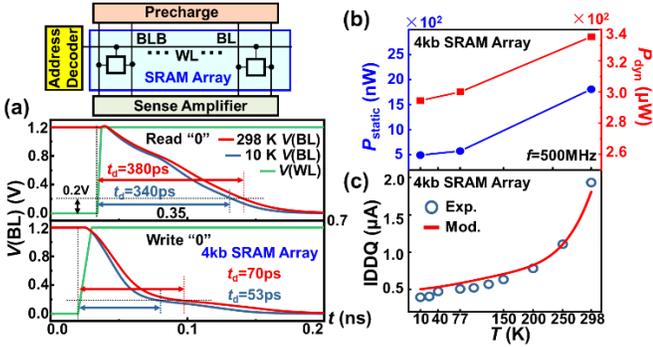

**Fig. 17.** (a) Read and write performance of the cryogenic SRAM array at $T$ = 298 K (red) and 10 K (blue). (b) Temperature-dependent static and dynamic power consumption and (c) the validation of the IDDQ current of the 4 Kb cryogenic SRAM circuit.

In the meantime, such a Cryo-CMOS model-assisted strategy proves to be effective for the cryogenic analog-to-digital converter (ADC) circuit design (*i.e.*, which serves as the key element to translate the analog readout signals from the quits to the digitalized ones). Here, since the threshold voltage invariably increases by around 0.2 V at low temperatures (Fig. 3), the conventional strong ARM-structure of the dynamic comparator module in room-temperature (RT)-ADC design fails to follow the bit switch when the bias level of the input differential pair is below 0.5 V with insufficient pull-down current. To address such challenge, we have re-designed the dynamic comparator unit by introducing an additional pre-amplifier stage to enlarge the common-mode voltage range and boost the input voltage swing under the guidance of the cryogenic PDK library. Consequently, the resulting 5-bit flash Cryo-ADC circuit can operate in a rail-to-rail mode at $T$ = 10 K with the sampling rate of 500 MSa/s, as highlighted by the real-time waveforms in Fig. 18.

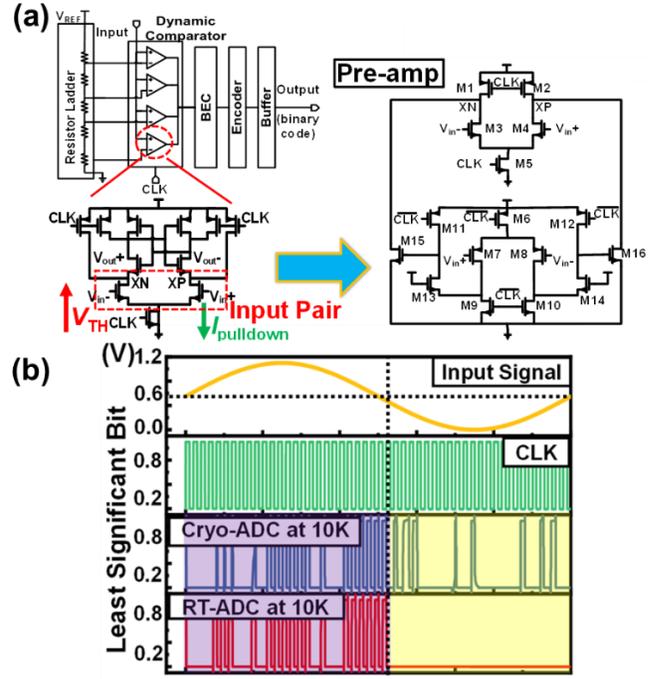

**Fig. 18** (a) Cryogenic 5-bit flash ADC circuit diagram and (b) post-layout simulation results to visualize the high-speed rail-to-rail operation at $T$ = 10 K.

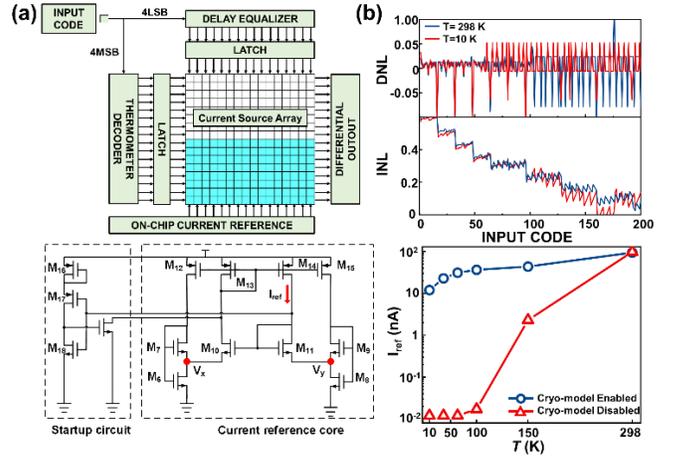

**Fig. 19** (a) Cryogenic 8-bit current steering DAC with an on-chip self-biased current reference module. (b) The DNL/INL and temperature-dependent $I_{ref}$ results validate the circuit performance. Data were re-captured from [33].

Additionally, this generic platform also enables the design of the cryogenic 8-bit current steering digital-to-analog converter (DAC) which bridges the digital instructions and the quantum processor driver. Due to the lack of a reliable Cryo-CMOS model, previous cryogenic DAC designs had to rely on external current references at room temperature, which inevitably introduced temperature variation-related errors and noises [32]. Alternatively, we have developed an on-chip self-biased current reference structure in which an additional inverter-type start-up circuit was used to ensure a well-defined quiescent operation point in the entire temperature region. Meanwhile, all the NMOS/PMOS sizes have been deliberately adjusted in accordance with the low-temperature device *I-V* behaviors so that both the circuit function and performance benchmark are well-maintained [33]. As emphasized in Fig. 19b, the simulated reference current $I_{ref}$ remains at the required nA-level from 298 K to 10 K, whereas the control circuit (*i.e.*, which was designed based on the room-temperature PDK) fails to produce sufficient reference current when $T < 100$ K. Moreover, by integrating this on-chip current reference module into the 8-bit cryogenic DAC system which adopted the standard 4 MSB + 4 LSB segmented structure, the overall post-layout results show an improved performance in terms of a compact chip area, sufficient spurious-free dynamic range (SFDR) of 58 dB, high sampling rate of 140 MS/s, and low power consumption of 13.8 μW [33].

## V. Conclusion

In conclusion, this work elaborates a generic cryogenic CMOS platform that encompasses device characterization, modeling, and EDA-guided circuit simulation. Both the temperature-dependent DC and RF measurements and the comprehensive variation/mismatch statistical analysis warrant the accountability of the proposed Cryo-BSIM device compact model. By extending the PDK library to the sub-10 K region, we demonstrate the effectiveness of the proposed bottom-up (*i.e.*, from the device level to the circuit-level) design strategy for the design and optimization of cryogenic analog/digital integrated circuits, thus providing a reliable framework for implementing high-performance cryogenic applications.